\documentclass[aps,prd,onecolumn,groupedaddress,showpacs,nofootinbib,amssymb]{revtex4}
\usepackage[dvips]{graphicx}
\usepackage{amssymb}
\usepackage{amsmath}
\usepackage{graphicx,color}
\usepackage{amsfonts}
\usepackage{bm}
\usepackage{cancel}
\usepackage{comment}

\newcommand\be{\begin{equation}}
\newcommand\ee{\end{equation}}

\allowdisplaybreaks[4]

\begin{document}

\tolerance=5000

\title{Neutron Stars Phenomenology with Scalar-tensor Inflationary Attractors}
\author{S.D.~Odintsov,$^{1,2}$\,\thanks{odintsov@ieec.uab.es}
V.K.~Oikonomou,$^{3,4}$\,\thanks{v.k.oikonomou1979@gmail.com}}
 \affiliation{$^{1)}$ ICREA, Passeig Luis Companys, 23, 08010 Barcelona, Spain\\
$^{2)}$ Institute of Space Sciences (IEEC-CSIC) C. Can Magrans
s/n,
08193 Barcelona, Spain\\
$^{3)}$ Department of Physics, Aristotle University of
Thessaloniki, Thessaloniki 54124,
Greece\\
$^{4)}$ Laboratory for Theoretical Cosmology, Tomsk State
University of Control Systems and Radioelectronics, 634050 Tomsk,
Russia (TUSUR)}

\tolerance=5000

\begin{abstract}
In this work we shall study the implications of a subclass of
$E$-models cosmological attractors, namely of $a$-attractors, on
hydrodynamically stable slowly rotating neutron stars.
Specifically, we shall present the Jordan frame theory of the
$a$-attractors, and by using a conformal transformation we shall
derive the Einstein frame theory. We discuss the inflationary
context of $a$-attractors in order to specify the allowed range of
values for the free parameters of the model based on the latest
cosmic-microwave-background-based Planck 2018 data. Accordingly,
using the notation and physical units frequently used in
theoretical astrophysics contexts, we shall derive the
Tolman-Oppenheimer-Volkoff equations in the Einstein frame.
Assuming a piecewise polytropic equation of state, the lowest
density part of which shall be chosen to be the WFF1, or APR or
the SLy EoS, we numerically solve the Tolman-Oppenheimer-Volkoff
equations using a double shooting python-based ``LSODA'' numerical
code. The resulting picture depends on the value of the parameter
$a$ characterizing the $a$-attractors. As we show, for large
values of $a$, which do not produce a viable inflationary era, the
$M-R$ graphs are nearly identical to the general relativistic
result, and these two are discriminated at large central densities
values. Also, for large $a$-values, the WFF1 equation of state is
excluded, due to the GW170817 constraints on the radius of an
$M\sim 1.6 M_{\odot}$ neutron star, which must be larger than
$R=10.68^{+15}_{-0.04}$km and on the radius corresponding to the
maximum mass which must be larger than $R=9.6^{+0.14}_{-0.03}$km.
In addition, the small $a$ cases produce larger masses and radii
compared to the general relativistic case and are compatible with
the GW170817 constraints on the radii of neutron stars. A notable
feature is that as the parameter $a$ decreases, the radii of the
static hydrodynamically stable neutron stars increase. Our results
indicate deep and not yet completely understood connections
between non-minimal inflationary attractors and neutron stars
phenomenology in scalar-tensor theory.
\end{abstract}


\maketitle

\section*{Introduction}

After the exciting results of the LIGO-Virgo collaboration
obtained in the last five years, neutron stars (NS) enjoy an
elevated role in the scientific community, since many scientific
theoretical frameworks can be tested by using neutron stars. For
example nuclear theory models
\cite{Tolos:2020aln,Lattimer:2012nd,Steiner:2011ft,Horowitz:2005zb,Watanabe:2000rj,Shen:1998gq,Xu:2009vi,Hebeler:2013nza,Mendoza-Temis:2014mja,Ho:2014pta,Kanakis-Pegios:2020kzp},
high energy particle models
\cite{Buschmann:2019pfp,Safdi:2018oeu,Hook:2018iia,Edwards:2020afl},
modified gravity models
\cite{Astashenok:2020qds,Capozziello:2015yza,Astashenok:2014nua,Astashenok:2014pua,Astashenok:2013vza,Arapoglu:2010rz,Astashenok:2020cqq}
can be tested, but also NS are in the epicenter of also
astrophysical models
\cite{Sedrakian:2015krq,Khadkikar:2021yrj,Sedrakian:2006zza,Sedrakian:2018kdm,Bauswein:2020kor,Vretinaris:2019spn,Bauswein:2020aag,Bauswein:2017vtn}.
The two most important observations coming from the LIGO-Virgo
collaboration are the GW170817 event
\cite{TheLIGOScientific:2017qsa} and the GW190814 event
\cite{Abbott:2020khf}, with the first indicating that the
gravitational wave speed is nearly equal to the speed of light,
and the latter indicating the possibility of either having an
astrophysical black hole or a NS with mass belonging to the
mass-gap region, although alternative scenarios can also be true
\cite{Biswas:2020xna}. For an important stream of textbooks and
reviews on NS we refer the reader to Refs.
\cite{Haensel:2007yy,Friedman:2013xza,Baym:2017whm,Lattimer:2004pg,Olmo:2019flu}.

Modified gravity in its various forms
\cite{Nojiri:2017ncd,Capozziello:2011et,Capozziello:2010zz,Nojiri:2006ri,
Nojiri:2010wj,delaCruzDombriz:2012xy,Olmo:2011uz,dimo}, enjoys an
elevated role among the various theoretical descriptions that can
harbor larger NS masses compared to general relativity (GR). The
most important and fundamental candidate modified gravity theory,
namely $f(R)$ gravity, can yield large mass NSs
\cite{Astashenok:2014nua}, and solve the hyperon puzzle for a
hyperon containing equation of state (EoS)
\cite{Astashenok:2014pua}. Apart from the applications in NSs,
modified gravity seems as an inevitable description for cosmology,
since dark energy cannot be consistently described solely by GR.
The reason is simple, the dark energy EoS $\omega_{DE}$ according
to the latest Planck cosmological constraints
\cite{Aghanim:2018eyx}, is constrained to have values
$\omega_{DE}=-1.018\pm 0.031$. So it is a probability, not just a
possibility, that dark energy is actually phantom dark energy.
Phantom dark energy in the context of GR is described only by
phantom scalars \cite{Caldwell:2003vq}, which do not provide a
physically acceptable description of nature. On the other hand,
crossing the phantom divide line in the context of modified
gravity is not an issue, since this can be achieved without
resorting to phantom scalar fields, see the reviews
\cite{Nojiri:2017ncd,Capozziello:2011et,Capozziello:2010zz,Nojiri:2006ri,
Nojiri:2010wj,delaCruzDombriz:2012xy,Olmo:2011uz} for details.

In view of the above, in this work we shall study static NSs in
hydrodynamic equilibrium in the context of scalar-tensor gravity.
Scalar-tensor studies of NSs already exist in the literature for
various theories, see for example
\cite{Pani:2014jra,Staykov:2014mwa,Horbatsch:2015bua,Silva:2014fca,Doneva:2013qva,Xu:2020vbs,Salgado:1998sg,Shibata:2013pra,Arapoglu:2019mun,Ramazanoglu:2016kul,AltahaMotahar:2019ekm,Chew:2019lsa,Blazquez-Salcedo:2020ibb,Motahar:2017blm},
but in this work we shall examine some $a$-attractor scalar models
\cite{alpha1,alpha2,alpha3,alpha4,alpha5,alpha6,alpha7,alpha8,alpha9,alpha10,alpha11,alpha12,alpha13,alpha14,alpha15,alpha16,alpha17,alpha18,alpha19,alpha20,alpha21,alpha22,alpha23,alpha24,alpha25,alpha26,alpha27,alpha28,alpha29,alpha30,alpha31,alpha32,alpha33,alpha34,alpha35,alpha36,alpha37}
which are unknown to the theoretical astrophysics community, but
are quite popular in theoretical cosmology contexts. This is due
to the fact that the cosmological attractors
\cite{alpha1,alpha2,alpha3,alpha4,alpha5,alpha6,alpha7,alpha8,alpha9,alpha10,alpha11,alpha12,alpha13,alpha14,alpha15,alpha16,alpha17,alpha18,alpha19,alpha20,alpha21,alpha22,alpha23,alpha24,alpha25,alpha26,alpha27,alpha28,alpha29,alpha30,alpha31,alpha32,alpha33,alpha34,alpha35,alpha36,alpha37}
have a main characteristic, they predict the same final form of
the spectral index of the primordial scalar curvature
perturbations and of the tensor-to-scalar ratio, for a wide class
of distinct scalar potentials and non-minimal coupling. More
importantly, the aforementioned indices are compatible in most
cases with the latest Planck constraints on inflation
\cite{Akrami:2018odb}. Our aim is to study in detail the Einstein
frame counterpart implications on static NS in hydrodynamic
equilibrium, of several classes of non-minimally coupled
$a$-attractor models in the Jordan frame. We shall present both
the inflationary theory, in order to better formulate the NS
study, using the cosmologist's perspective, and then derive the
Einstein frame Tolman-Oppenheimer-Volkoff (TOV) equations using
Geometrized units and by adopting the usual notation of
theoretical astrophysics contexts. With regard to the EoS, we
shall use a piecewise polytropic EoS
\cite{Read:2008iy,Read:2009yp}, and in our study we shall extract
from the numerical results the Jordan frame Arnowitt-Deser-Misner
(ADM) mass \cite{Arnowitt:1960zzc}, which for static NSs coincides
\cite{Shibata:2013ssa} with the Komar mass \cite{Komar:1958wp}.
The numerical analysis will be based on a modified version of the
freely available python 3 based numerical code \cite{niksterg},
with the TOV equations changed in order to align the code with
Ref. \cite{Pani:2014jra}, and also the Jordan frame radii and
masses of the NSs are calculated, not just the Einstein frame
ones.

\section{$a$-attractors Inflation and Theoretical Framework}

Before we present the TOV equations in the Einstein frame, using
the standard notation used in theoretical astrophysics context,
let us first recall the theoretical framework of inflationary
dynamics of $a$-attractor potentials, see Refs.
\cite{alpha1,alpha2,alpha3,alpha4,alpha5,alpha6,alpha7,alpha8,alpha9,alpha10,alpha11,alpha12,alpha13,alpha14,alpha15,alpha16,alpha17,alpha18,alpha19,alpha20,alpha21,alpha22,alpha23,alpha24,alpha25,alpha26,alpha27,alpha28,alpha29,alpha30,alpha31,alpha32,alpha33,alpha34,alpha35,alpha36,alpha37}
for details. Also in order to render the article self-contained,
we shall recall to some limited extent the conformal
transformation notation from the cosmologist perspective, see
\cite{Kaiser:1994vs,valerio,Faraoni:2013igs,Buck:2010sv} for more
details.

The Jordan frame action of non-minimally coupled scalar field in
the presence of perfect matter fluids is,
\begin{equation}\label{c1}
\mathcal{S}_J=\int
d^4x\Big{[}f(\phi)R-\frac{\omega(\phi)}{2}g^{\mu
\nu}\partial_{\mu}\phi\partial_{\nu}\phi-U(\phi)\Big{]}+S_m(g_{\mu
\nu},\psi_m)\, ,
\end{equation}
with $\psi_m$ being the Jordan frame perfect matter fluids present
in the Jordan frame, which have pressure $P$ and energy density
$\epsilon$. For this section we shall use natural units, however
for the NS study we shall use Geometrized units, in order to
comply with the NS literature and notation. In addition, the
Jordan frame metric is $g_{\mu \nu}$ and the minimal coupling case
in the action (\ref{c1}) corresponds to,
\begin{equation}\label{c2}
f(\phi)=\frac{1}{16 \pi G}=\frac{M_p^2}{2}\, ,
\end{equation}
with,
\begin{equation}\label{c3}
M_p=\frac{1}{\sqrt{8\pi G}}\, ,
\end{equation}
which is the Jordan frame reduced Planck mass, while $G$ is
Newton's gravitational constant in the Jordan frame. Upon
performing the following conformal transformation,
\begin{equation}\label{c4}
\tilde{g}_{\mu \nu}=\Omega^2g_{\mu \nu}\, ,
\end{equation}
the Einstein frame action is obtained, with the tilde denoting
physical quantities in the Einstein frame. The minimal scalar
field action in the Einstein frame is obtained by making the
choice, \cite{Kaiser:1994vs,valerio},
\begin{equation}\label{c6}
\Omega^2=\frac{2}{M_p^2}f(\phi)\, ,
\end{equation}
hence upon performing the conformal transformation for all the
quantities in the action (\ref{c1}) and for the choice (\ref{c6}),
one obtains the Einstein frame action,
\begin{equation}\label{c12}
\mathcal{S}_E=\int
d^4x\sqrt{-\tilde{g}}\Big{[}\frac{M_p^2}{2}\tilde{R}-\frac{\zeta
(\phi)}{2} \tilde{g}^{\mu \nu }\tilde{\partial}_{\mu}\phi
\tilde{\partial}_{\nu}\phi-V(\phi)\Big{]}+S_m(\Omega^{-2}\tilde{g}_{\mu
\nu},\psi_m)\, ,
\end{equation}
with the Einstein frame scalar potential $V(\phi)$ being related
to the Jordan frame one $U(\phi)$ as,
\begin{equation}\label{c13}
V(\phi)=\frac{U(\phi)}{\Omega^4}\, ,
\end{equation}
and also $\zeta(\phi)$ is defined as follows,
\begin{equation}\label{c14}
\zeta
(\phi)=\frac{M_p^2}{2}\Big{(}\frac{3\Big{(}\frac{df}{d\phi}\Big{)}^2}{f^2}+\frac{2\omega(\phi)}{f}\Big{)}\,
.
\end{equation}
We can make the scalar field kinetic term $\zeta(\phi)$ canonical,
upon making the following transformation,
\begin{equation}\label{c15}
\Big{(}\frac{d\varphi}{d \phi}\Big{)} =\sqrt{\zeta(\phi)}\, ,
\end{equation}
and in effect, the Einstein frame canonical scalar field action
reads,
\begin{equation}\label{c17}
\mathcal{S}_E=\int
d^4x\sqrt{-\tilde{g}}\Big{[}\frac{M_p^2}{2}\tilde{R}-\frac{1}{2}\tilde{g}^{\mu
\nu } \tilde{\partial}_{\mu}\varphi
\tilde{\partial}_{\nu}\varphi-V(\varphi)\Big{]}+S_m(\Omega^2\tilde{g}_{\mu
\nu},\psi_m)
\end{equation}
where,
\begin{equation}\label{c18}
V(\varphi)=\frac{U(\varphi)}{\Omega^4}=\frac{U(\varphi)}{4
M_p^4f^2}\, .
\end{equation}
It is important to note that in the Einstein frame, the matter
fluids are not perfect, since the energy momentum tensor
satisfies,
\begin{equation}\label{c24}
\tilde{\partial}^{\mu}\tilde{T}_{\mu \nu}=-\frac{d}{d\varphi}[\ln
\Omega]\tilde{T}\tilde{\partial}_{\nu}\phi\, ,
\end{equation}
where. Also, the pressure and the energy density are transformed
from the Einstein to the Jordan frame as follows,
\begin{equation}\label{c28}
\tilde{\varepsilon}=\Omega^{-4}(\varphi)\varepsilon,\,\,\,\tilde{P}=\Omega^{-4}(\varphi)P\,
.
\end{equation}
Let us now discuss how the $a$-attractor and the subclass of
$E$-model attractor models are obtained. The $a$-attractor
potentials can be obtained if we choose,
\begin{equation}\label{alphaattrcondition}
\omega (\phi)=\frac{\Big{(}\frac{df}{d\phi}\Big{)}^2}{4\xi f}\, ,
\end{equation}
where $\xi$ is an arbitrary parameter of the model. By using the
choice (\ref{alphaattrcondition}), the relation (\ref{c15})
becomes,
\begin{equation}\label{c15alpha}
\Big{(}\frac{d\varphi}{d \phi}\Big{)}
=M_p\sqrt{\frac{3a}{2}}\frac{\Big{(}\frac{df}{d\phi}\Big{)}}{ f}\,
,
\end{equation}
where we introduced the parameter $a$, defined as follows,
\begin{equation}\label{alphaprm}
a=1+\frac{1}{6\xi}\, .
\end{equation}
Obviously, values of $a$ less than unity correspond to theories
with negative $\xi$, while values of $a$ larger than unity
correspond to positive $\xi$ values. In view of Eq. (\ref{c15})
and for the choice (\ref{c15alpha}), we get,
\begin{equation}\label{c15newalpha}
\varphi=M_p\sqrt{\frac{3\alpha}{2}}\ln f\, ,
\end{equation}
and notice that the relation above holds true irrespective the
choice of the arbitrary function $f$. Already the unique
universality of the $a$-attractor potentials makes its presence
apparent. From Eq. (\ref{c15newalpha}), we get,
\begin{equation}\label{fasfunctionofphi}
f=e^{\sqrt{\frac{2}{3\alpha }}\frac{\varphi}{M_p}}\, .
\end{equation}
Now the $E$-models attractors and the $a$-attractors are obtained
if the Jordan frame potential is generally chosen as,
\begin{equation}\label{alphaattractorsjordanpot}
U(\phi)=V_0f^2\Big{(}1-\frac{1}{f}\Big{)}^{2n}\, ,
\end{equation}
thus the Einstein frame potential reads,
\begin{equation}\label{einsteinframepot}
V(\varphi)=\mathcal{V}_0\Big{(}1-\frac{1}{f}\Big{)}^{2n}=\tilde{V}_0M_p^4\Big{(}1-\frac{1}{f}\Big{)}^{2n}\,
,
\end{equation}
where $\mathcal{V}_0=\tilde{V}_0M_p^4$ and
$\tilde{V}_0=\frac{V_0}{4}$, thus $\tilde{V}_0$ is dimensionless
in natural units and $\mathcal{V}_0$ has dimensions $[m]^4$. The
class of potentials (\ref{alphaattractorsjordanpot}) for $f$ given
in Eq. (\ref{fasfunctionofphi}), is known as $E$-models potentials
while for $n=2$ the much more popular $a$-attractor potentials are
obtained. For all $a$ values both the $a$ and $E$-attractor
potentials (\ref{einsteinframepot}) yield the same spectral index
of the primordial scalar curvature perturbations $n_s$ and the
same tensor-to-scalar ratio $r$ \cite{alpha3}. We shall be
interested in values of $a$ less than unity (the values can be
extended to be of order $\mathcal{O}(10)$ and the theory can still
be valid, see for example Figure 1 of Ref. \cite{alpha3}, but we
choose a safe value for $a$, namely $a=0.6$), and much larger than
$\frac{8N}{3}$, where $N$ is the $e$-foldings number which for
successful inflation takes the values $N\sim 50-60$. Specifically
for the NS study we shall choose $a=0.6$ and $a=10^4$, so let us
discuss the inflationary dynamics of the two different limiting
cases of $a$. As it was shown in \cite{alpha3}, for $a$ being
small or of the order of unity, the spectral index and the
tensor-to-scalar ratio at leading order in $N$ are,
\begin{equation}\label{spectralindexsmallalpha}
n_s\simeq 1-\frac{2}{N}\, ,\,\,\,r=\frac{12a}{N^2}\, .
\end{equation}
Thus, regardless the choice of the function $f(\phi(\varphi))$,
all the models (\ref{einsteinframepot}) yield the same
inflationary dynamics, and this justifies the terminology
attractors. In this range, the dependence of the resulting
phenomenology on $n$ is practically negligible. As we already
mentioned, we shall focus on the case $n=1$, so we shall be
interested in the $a$-attractors. The analysis of the NSs
implications for general $n$ is similar, so we shall not discuss
this case in this paper. The inflationary phenomenology implied by
the observational indices of Eq. (\ref{spectralindexsmallalpha})
is viable and compatible with the latest Planck data for a wide
range of the parameter $a$, and certainly for $a$ of the order of
unity (see Fig. 1 of Ref. \cite{alpha3} for details). Of course
for $a=0.6$ the inflationary phenomenology is viable and the
tensor-to-scalar ratio is smaller compared to the $R^2$ model.
Also for $a\gg \frac{8N}{3}$, the spectral index and the
tensor-to-scalar ratio read \cite{alpha3},
\begin{equation}\label{spectralindexinfalpha}
n_s=1-\frac{2}{N},\,\,\, r=\frac{8}{N}\, .
\end{equation}
Note that the observational indices of inflation of Eq.
(\ref{spectralindexinfalpha}) cannot be compatible with the Planck
data \cite{Akrami:2018odb} for $N\sim 50-60$, but we shall include
this case in our NS study of the next section, since the potential
(\ref{einsteinframepot}) provides phenomenology which basically
interpolates between the cases described by Eqs.
(\ref{spectralindexsmallalpha}) and (\ref{spectralindexinfalpha}).
Interestingly enough, the limiting case $a\sim\mathcal{O}(1)$
which yields a viable inflationary phenomenology, results to NSs
which deviate significantly from the GR case, while the large $a$
case results to NSs which mimic to some extent the GR behavior.

Before closing, let us demonstrate which values of the parameter
$\tilde{V}_0$ defined below Eq. (\ref{einsteinframepot}) can be
compatible with the latest Planck constraints. According to the
latest Planck data, the amplitude of the scalar fluctuations for
canonical single field scalar models $\Delta_s^2$ defined as,
\begin{equation}\label{scalaramp}
\Delta_s^2=\frac{1}{24\pi^2}\frac{V(\varphi_f)}{M_p^4}\frac{1}{\epsilon(\varphi_f)}\,
,
\end{equation}
is constrained to be \cite{Akrami:2018odb},
\begin{equation}\label{scalarampconst}
\Delta_s^2=2.2\times 10^{-9}\, .
\end{equation}
In Eq. (\ref{scalaramp}), $\varphi_f$ is the value of the scalar
field at the end of inflation, and also $\epsilon$ is the first
slow-roll index. This means that in our case,
\begin{equation}\label{tilde}
\mathcal{V}_0=\tilde{V}_0M_p^4\sim 9.6\times 10^{-11}\, M_p^4\, .
\end{equation}
The above constraint shall be taken into account for the NSs
study, because the parameter $\mathcal{V}_0$ cannot be arbitrary,
and of course we shall transform the constraint (\ref{tilde}) in
Geometrized units and not natural units.

Before continuing, let us quote here a useful expression for the
inflationary action in the Einstein frame,
\begin{equation}\label{einsteinframeaction}
\mathcal{S}_E=\int
d^4x\sqrt{-\tilde{g}}\Big{[}\frac{M_p^2}{2}\tilde{R}-\frac{1}{2}\tilde{g}^{\mu
\nu } \tilde{\partial}_{\mu}\varphi
\tilde{\partial}_{\nu}\varphi-V(\varphi)\Big{]}\, ,
\end{equation}
which can be rewritten as,
\begin{equation}\label{einsteinframeactioninflationns}
\mathcal{S}_E=\int d^4x\sqrt{-\tilde{g}}\Big{[}\frac{1}{16\pi
G}\tilde{R}-\frac{1}{2}\tilde{g}^{\mu \nu }
\tilde{\partial}_{\mu}\varphi
\tilde{\partial}_{\nu}\varphi-\frac{16\pi G V(\varphi)}{16\pi
G}\Big{]}\, ,
\end{equation}
where recall that $M_p^2=\frac{1}{8\pi G}$. We shall make use of
Eq. (\ref{einsteinframeactioninflationns}) for our next section NS
study.

\section{Neutron Stars in the Einstein Frame with $a$-attractors Scalar Potential}

Now we proceed to the study of NSs for $a$-attractor potentials we
discussed in the previous section. We shall adopt though a
different notation from the previous section in order to align the
presentation with the theoretical astrophysics works, so we use
the notation of Ref. \cite{Pani:2014jra} with the only difference
being the use of tilde for the Einstein frame quantities. The
general Jordan frame starting action of the non-minimally coupled
scalar field in Geometrized units ($G=c=1$) is
\cite{Pani:2014jra},
\begin{equation}\label{ta}
\mathcal{S}=\int
d^4x\frac{\sqrt{-g}}{16\pi}\Big{[}f(\phi)R-\omega(\phi)g^{\mu
\nu}\partial_{\mu}\phi\partial_{\nu}\phi-U(\phi)\Big{]}+S_m(\psi_m,g_{\mu
\nu})\, .
\end{equation}
By performing the following conformal transformation,
\begin{equation}\label{ta1higgs}
\tilde{g}_{\mu \nu}=A^{-2}g_{\mu \nu}\, ,
\end{equation}
and by choosing,
\begin{equation}\label{ta2higgs}
A(\phi)=f^{-1/2}(\phi)\, ,
\end{equation}
we may obtain the Einstein frame action, which may be expressed in
terms of a canonical scalar field $\varphi$ as follows,
\begin{equation}\label{ta5higgs}
\mathcal{S}=\int
d^4x\sqrt{-\tilde{g}}\Big{(}\frac{\tilde{R}}{16\pi}-\frac{1}{2}
\tilde{g}_{\mu \nu}\partial^{\mu}\varphi
\partial^{\nu}\varphi-\frac{V(\varphi)}{16\pi}\Big{)}+S_m(\psi_m,A^2(\varphi)g_{\mu
\nu})\, ,
\end{equation}
with the scalar field $\phi$ and the Einstein frame scalar field
being related as follows,
\begin{equation}\label{ta4higgs}
\frac{d \varphi }{d \phi}=\frac{1}{\sqrt{4\pi}}
\sqrt{\Big{(}\frac{3}{4}\frac{1}{f^2}\Big{(}\frac{d
f}{d\phi}\Big{)}^2+\frac{\omega(\phi)}{2f}\Big{)}}\, .
\end{equation}
The $a$-attractors potentials can be obtained by choosing,
\begin{equation}\label{omegaphi}
\omega (\phi)=\frac{1}{4\xi}\frac{1}{f}\Big{(}\frac{d
f}{d\phi}\Big{)}^2\, ,
\end{equation}
and also for the Jordan frame potential,
\begin{equation}\label{alphaattractpotens}
U(\phi)=\mathcal{U}_0f^2\left(1-\frac{1}{f} \right)^{2n}\, .
\end{equation}
By using Eq. (\ref{omegaphi}), Eq. (\ref{ta4higgs}) yields,
\begin{equation}\label{finalvarphiphi}
\frac{d \varphi }{d
\phi}=\sqrt{\frac{3a}{16\pi}}\frac{1}{f}\Big{(}\frac{d
f}{d\phi}\Big{)}\, ,
\end{equation}
where again $a$ is defined in Eq. (\ref{alphaprm}). Thus, by
integrating (\ref{finalvarphiphi}) we get,
\begin{equation}\label{fvarphi}
f=e^{\sqrt{\frac{16\pi}{3a}}\varphi}\, ,
\end{equation}
and holds true regardless the choice of the function $f$. The
Jordan frame potential is obtained by using the following
relation,
\begin{equation}\label{antegeia}
V(\varphi)=\frac{U(\phi)}{f^2}\, ,
\end{equation}
hence by combining Eqs. (\ref{alphaattractpotens}) and
(\ref{fvarphi}), we obtain the final form of the Einstein frame
potential which is,
\begin{equation}\label{einsteinframepotentialfinal}
V(\varphi)=\mathcal{U}_0\left(1-e^{-\sqrt{\frac{16\pi}{3a}}\varphi}
\right)^{2n}\, ,
\end{equation}
which will be the final form of the $a$-attractors potential in
the Einstein frame in Geometrized units. One can easily observe
that by putting $M_p=1/\sqrt{8\pi G}$ in Eq.
(\ref{fasfunctionofphi}), relation (\ref{fasfunctionofphi}) is
identical with Eq. (\ref{fvarphi}), and the same applies for the
potential. As a function of $\varphi$, the conformal factor
$A(\phi)$ reads,
\begin{equation}\label{Aofpvarphiprofinal}
A(\varphi)=e^{\alpha \varphi}\, ,
\end{equation}
where $\alpha$ stands for,
\begin{equation}\label{alphaofphi}
\alpha=-\frac{1}{2}\sqrt{\frac{16\pi}{3 a}}\, ,
\end{equation}
and in practise, $\alpha(\varphi)$ has the following definition in
general,
\begin{equation}\label{alphaofvarphigeneraldef}
\alpha(\varphi)=\frac{d A(\varphi)}{d \varphi}\, ,
\end{equation}
hence in the case at hand,
\begin{equation}\label{alphaofphifinalintermsofvarphi}
a(\varphi)=\alpha=-\frac{1}{2}\sqrt{\frac{16\pi}{3 a}}\, .
\end{equation}
In the same way, the scalar potential in the Einstein frame is
expressed in terms of the parameter $\alpha$ as follows,
\begin{equation}\label{einsteinframepotentialfinal}
V(\varphi)=\mathcal{U}_0\left(1-e^{2\alpha \varphi} \right)^{2n}\,
.
\end{equation}
\begin{figure}[h!]
\centering
\includegraphics[width=20pc]{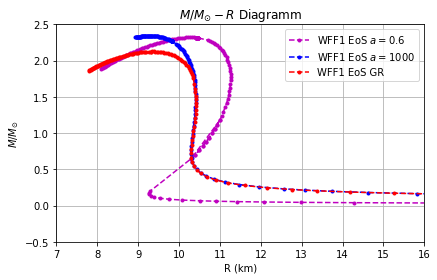}
\includegraphics[width=20pc]{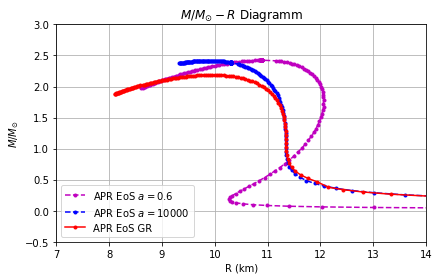}
\includegraphics[width=20pc]{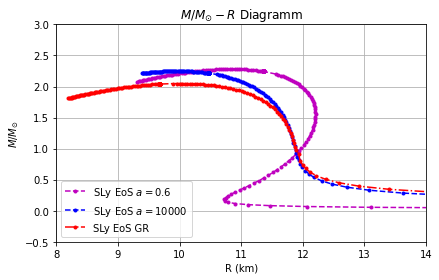}
\caption{$M-R$ graphs for the $a$-attractors model compared to the
GR case, for the WFF1 EoS (upper left plot), the APR EoS (upper
right plot) and the SLy EoS (bottom plot). The $y$-axis
corresponds to $M/M_{\odot}$, where $M$ is the Jordan frame ADM
mass of the neutron star, and the $x$-axis corresponds to the
Jordan frame circumferential radius of the neutron star expressed
in kilometers. In all the plots we took
$\mathcal{U}_0=7.62094\times 10^{-12}$, and the values of $a$ are
displayed in each plot.} \label{plot1}
\end{figure}
Now let us determine the values of $\mathcal{U}_0$ by making a
direct correspondence to the inflationary theory of the previous
section. Using Eq. (\ref{einsteinframeactioninflationns}) and Eq.
(\ref{tilde}) a direct consequence is that $\mathcal{U}_0$ in Eq.
(\ref{einsteinframepotentialfinal}) is $\mathcal{U}_0=16\pi
\mathcal{V}_0$ and since $M_p=1/\sqrt{8 \pi}$ in Geometrized
units, we get,
\begin{equation}\label{U0constraint}
\mathcal{U}_0=7.62094\times 10^{-12}\, ,
\end{equation}
and this is the value for the scalar potential parameter
$\mathcal{U}_0$ in Geometrized units. Now we proceed to the TOV
equations for the framework at hand, and we shall consider a
spherically symmetric and static metric of the form,
\begin{equation}\label{tov1}
ds^2=-e^{\nu(r)}dt^2+\frac{dr^2}{1-\frac{2
m}{r}}+r^2(d\theta^2+\sin^2\theta d\phi^2)\, ,
\end{equation}
with the function $m(r)$ describing the gravitational mass of the
neutron star of circumferential radius $r$.
\begin{figure}[h!]
\centering
\includegraphics[width=20pc]{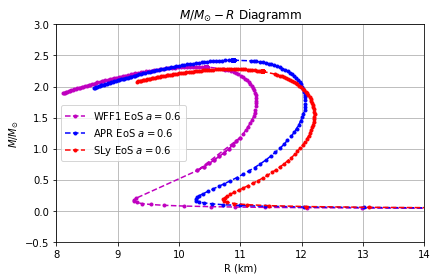}
\includegraphics[width=20pc]{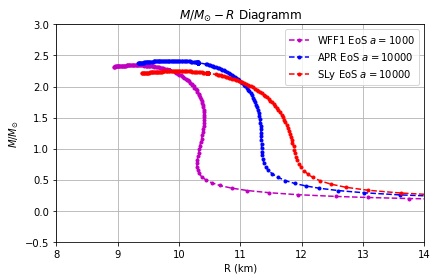}
\caption{Comparison of the $M-R$ graphs for the $a$-attractors
model for each value of the parameter $a$ for all the EoSs. In the
left plot we present the $M-R$ graphs for the WFF1, APR and SLy
EoSs for $a=0.6$ and in the right plot for $a=1000$ for the WFF1
EoS, and for $a=10000$ for the SLy and APR EoSs.} \label{plot2}
\end{figure}
For the metric (\ref{tov1}), the TOV equations are
\cite{Pani:2014jra},
\begin{equation}\label{tov2}
\frac{d m}{dr}=4\pi r^2
A^4(\varphi)\epsilon+\frac{r}{2}(r-2m)\omega^2+4\pi
r^2V(\varphi)\, ,
\end{equation}
\begin{equation}\label{tov3}
\frac{d\nu}{dr}=r\omega^2+\frac{2}{r(r-2m)}\Big{[}4\pi
A^4(\varphi)r^3P-4\pi V(\varphi) r^3\Big{]}+\frac{2m}{r(r-2m)}\, ,
\end{equation}
\begin{equation}\label{tov4}
\frac{d\omega}{dr}=\frac{4\pi r
A^4(\varphi)}{r-2m}\Big{(}\alpha(\varphi)(\epsilon-3P)+
r\omega(\epsilon-P)\Big{)}-\frac{2\omega
(r-m)}{r(r-2m)}+\frac{8\pi \omega r^2 V(\varphi)+r\frac{d
V(\varphi)}{d \varphi}}{r-2 m}\, ,
\end{equation}
\begin{equation}\label{tov5}
\frac{dP}{dr}=-(\epsilon+P)\Big{[}\frac{1}{2}\frac{d\nu}{dr}+\alpha
(\varphi)\omega\Big{]}\, ,
\end{equation}
where we defined the function $\alpha (\varphi)$ in Eqs.
(\ref{alphaofvarphigeneraldef}) and
(\ref{alphaofphifinalintermsofvarphi}) for the $a$-attractor
model. Furthermore, there is an additional equation to be
considered along with the other differential equations above,
namely $\omega=\frac{d \varphi}{dr}$. Moreover,  the pressure $P$
and energy density $\epsilon$ appearing in the TOV equations
above, are basically the Jordan frame quantities, which are
related to their Einstein frame counterparts as follows,
\begin{equation}\label{ta11}
\tilde{\epsilon}=A^4\epsilon \, ,
\end{equation}
\begin{equation}\label{ta12}
\tilde{P}=A^4 P \, .
\end{equation}
So the purpose of this section is to numerically solve in a
consistent way the TOV equations, and to this end we shall
consider the following initial conditions,
\begin{equation}\label{tov8}
P(0)=P_c\, ,
\end{equation}
\begin{equation}\label{tov9}
m(0)=0\, ,
\end{equation}
\begin{equation}\label{tov10}
\nu(0)\, ,=-\nu_c
\end{equation}
\begin{equation}\label{tov11}
\varphi(0)=\varphi_c\, ,
\end{equation}
\begin{equation}\label{tov12}
\omega (0)=0\, ,
\end{equation}
and we shall be choosing three different types of EoSs for the
neutron star nuclear matter. Specifically we shall consider
piecewise polytropic EoSs \cite{Read:2008iy,Read:2009yp} (see also
the text in \cite{niksterg}), in which the low-density part
corresponds to the WFF1, APR or the SLy EoS. It is worth
describing in brief the polytropic EoS in order to maintain the
article self-contained. A piecewise polytropic EoS is constructed
by a low-density part $\rho<\rho_0$, which is chosen to be a
tabulated crust EoS, and by a high density part $\rho\gg \rho_0$.
The density $\rho_0$ is obtained by matching the low and high
density pieces. The piecewise polytropic EoS further consists of
other two dividing high densities, $\rho_1 = 10^{14.7}{\rm
g/cm^3}$ and $\rho_2= 10^{15.0}{\rm g/cm^3}$, and for each of
these densities, the intermediate densities and pressures
$\rho_{i-1} \leq \rho \leq \rho_i$  satisfy,
\begin{equation}\label{pp1}
P = K_i\rho^{\Gamma_i}\, ,
\end{equation}
with the requirement of continuity in each of the crossing points
of each piece,  as follows,
\begin{equation}\label{pp2}
P(\rho_i) = K_i\rho^{\Gamma_i} = K_{i+1}\rho^{\Gamma_{i+1}}\, .
\end{equation}
Due to the continuity at each piece of the piecewise polytropic
EoS, by using Eq. (\ref{pp2}) one determines the parameters $K_2$
and $K_3$ for a given chosen values of the parameters $K_1,
\Gamma_1, \Gamma_2, \Gamma_3$, or basically for a given initial
pressure $p_1$ and for given parameters $\Gamma_2$, and
$\Gamma_3$. In this work we shall consider three types of EoS, the
 a variational method EoS the WFF1 \cite{Wiringa:1988tp}, a potential method EoS the SLy
\cite{Douchin:2001sv}, and the APR EoS \cite{Akmal:1998cf}, so the
parameters $p_1$ and the parameters $\Gamma_2$, and $\Gamma_3$
will be chosen according to the values determined by the above
EoSs. Now let us quote the relation of the energy density as a
function of the pressure for the piecewise polytropic EoS, which
can be derived by simply integrating the first law of
thermodynamics,
\begin{equation}\label{pp3}
d\frac{\epsilon}{\rho} = - P d\frac{1}{\rho}\, ,
\end{equation}
for a barotropic fluid, and by using the continuity of the energy
density we get,
\begin{equation}\label{pp4}
\epsilon(\rho) = (1+\alpha_i)\rho +
\frac{K_i}{\Gamma_i-1}\rho^{\Gamma_i}\, ,
\end{equation}
which holds true for $\Gamma_i \neq 1$, and also $\alpha_i$ is,
\begin{equation}\label{pp5}
\alpha_i = \frac{\epsilon(\rho_{i-1})}{\rho_{i-1}} -1 -
\frac{K_i}{\Gamma_i-1}\rho_{i-1}^{\Gamma_i-1}\, .
\end{equation}
Before integrating the TOV equations, we need to clarify the mass
of the neutron star issue, and we shall use the ADM mass. The
numerical integration we shall perform on the TOV equations, shall
yield the Einstein frame ADM mass, but we need to extract the
Jordan frame ADM mass of the NS, since it is the Jordan frame mass
which can be measured in observation via observing the Keplerian
weak field orbits. Let us proceed to express the Jordan frame mass
in terms of the Einstein frame mass and in terms of the scalar
field. Let $h_E$ and $h_J$ be defined as follows in Geometrized
units,
\begin{equation}\label{hE}
\mathcal{K}_E=1-\frac{2 m}{r_E}\, ,
\end{equation}
\begin{equation}\label{hE}
\mathcal{K}_J=1-\frac{2  m_J}{r_J}\, .
\end{equation}
The relation between $\mathcal{K}_E$ and $\mathcal{K}_J$ in the
two frames is,
\begin{equation}\label{hehjrelation}
\mathcal{K}_J=A^{-2}\mathcal{K}_E\, ,
\end{equation}
and furthermore the radii of the NS in the two frames satisfy,
\begin{equation}\label{radiiconftrans}
r_J=A r_E\, ,
\end{equation}
and the Jordan frame ADM mass is defined as follows,
\begin{equation}\label{jordaframemass1}
M_J=\lim_{r\to \infty}\frac{r_J}{2}\left(1-\mathcal{K}_J \right)
\, ,
\end{equation}
while the Einstein frame ADM mass is defined as follows,
\begin{equation}\label{einsteiframemass1}
M_E=\lim_{r\to \infty}\frac{r_E}{2}\left(1-\mathcal{K}_E \right)
\, .
\end{equation}
If the asymptotic limit of Eq. (\ref{hehjrelation}) is taken, we
have,
\begin{equation}\label{asymptotich}
\mathcal{K}_J(r_E)=\left(1+\alpha(\varphi(r_E))\frac{d \varphi}{d
r}r_E \right)^2\mathcal{K}_E(\varphi(r_E))\, ,
\end{equation}
and hereafter $r_E$ will stand for the Einstein frame radius at
large distances from the surface of the NS, near the numerical
infinity but not exactly at the numerical infinity, and also
$\frac{d\varphi }{dr}=\frac{d\varphi }{dr}\Big{|}_{r=r_E}$. By
combining Eqs. (\ref{hE})-(\ref{asymptotich}) we finally get,
\begin{equation}\label{jordanframeADMmassfinal}
M_J=A(\varphi(r_E))\left(M_E-\frac{r_E^{2}}{2}\alpha
(\varphi(r_E))\frac{d\varphi
}{dr}\left(2+\alpha(\varphi(r_E))r_E\frac{d \varphi}{dr}
\right)\left(1-\frac{2 M_E}{r_E} \right) \right)\, ,
\end{equation}
and recall that $\frac{d\varphi }{dr}=\frac{d\varphi
}{dr}\Big{|}_{r=r_E}$. For our numerical analysis, after
calculating the Einstein frame mass, we shall use the relation
(\ref{jordanframeADMmassfinal}) in order to extract the Jordan
frame mass, and also with $M$ we will denote the Jordan frame mass
$M_J$, namely $M=M_J$.

Turning our focus to the neutron star radius, the Einstein frame
one $R_s$ is determined by the condition $P(R_s)=0$, thus the
Jordan frame radius of the NS, namely $R$, will be calculated by
using the following formula,
\begin{equation}\label{radiussurface}
R=A(\varphi(R_s))\, R_s\, ,
\end{equation}
and the result will be expressed in kilometers. Our aim in this
section is to numerically solve the TOV equations for the interior
and the exterior of the NS (where $\epsilon=P=0$ and the scalar
potential is non-zero), and we shall extract the Jordan frame mass
and radius of the NS and construct the corresponding $M-R$ graph.
The numerical code we shall use is a modified version of the
python 3 based code pyTOV-STT \cite{niksterg}, using the ``LSODA''
numerical integrator, for which the Jordan frame mass and radius
will be calculated. The numerical analysis will include a double
shooting method in order to extract the optimal values of $\nu_c$
and $\varphi_c$ at the center of the NS, which provide a
sufficient decay of the values of the Einstein frame scalar field
and of the metric function $\nu(r)$  at the numerical infinity,
which is chosen to be $r\sim 67.94378528694695$ km in the Einstein
frame. As we already mentioned, we shall use piecewise polytropic
EoSs, with the parameters $p_1$, $\Gamma_2$, and $\Gamma_3$
corresponding to the WFF1 \cite{Wiringa:1988tp}, the SLy
\cite{Douchin:2001sv}, and the APR EoS \cite{Akmal:1998cf}.
Finally, we shall study the cases for which $a=0.6$ and $a=10000$
for the APR and SLy, and also the cases $a=0.6$ and $a=1000$ for
the WFF1.

The results of our numerical analysis can be seen in Figs.
\ref{plot1} and \ref{plot2}, and now we shall critically examine
the results of our analysis. In Fig. \ref{plot1} we present the
$M-R$ graphs for the $a$-attractors model compared to the GR case,
for the WFF1 EoS (upper left plots), the APR EoS (upper right
plots) and the SLy EoS (bottom plots). The $y$-axis corresponds to
$M/M_{\odot}$, where $M$ is the Jordan frame ADM mass of the
neutron star, and the $x$-axis corresponds to the Jordan frame
circumferential radius of the neutron star $R$ expressed in
kilometers. In all the plots we took $\mathcal{U}_0=7.62094\times
10^{-12}$, and the values of $a$ are displayed in each plot. As it
can be seen from all the plots, the large $a$ curves mimic
significantly the corresponding GR curve, for a large number of
central densities, and deviate from the GR curve from a certain
point of high central density. Readily, the large $a$ case for the
WFF1 EoS is excluded from being a viable NS configuration, since
the GW170817 event constrains neutron stars with masses $M\sim 1.6
M_{\odot}$ area, to have radii in the range
$R=10.68^{+15}_{-0.04}$km \cite{Bauswein:2017vtn}. Thus only for
the WFF1 EoS, the large $a$ values are excluded, and it is
remarkably interesting that for large $a$ values, the
corresponding inflationary theory is not compatible with the
Planck constraints on inflation. On the other hand, the small $a$
case for the WFF1 EoS is compatible with the aforementioned
constraint of the GW170817 event. A common feature of all the
$M-R$ curves for $a$ values near unity is that the stable NS
configurations have larger radii, and specifically, as the values
of $a$ decrease, the radii of the NSs increase. This can be seen
from all the plots of Fig. \ref{plot1}, but also in Table
\ref{table1} where we present the maximum masses for the
$a$-attractor models and the correspondent radii. From Table
\ref{table1}, in conjunction with Fig. \ref{plot1}, we can also
verify that the GW170817 constraint \cite{Bauswein:2017vtn}, which
constrains the radii of the maximum mass static configurations to
be larger than $R=9.6^{+0.14}_{-0.03}$km, is satisfied for all the
cases studied, except for $a=1000$ for the WFF1 EoS which is
excluded.
\begin{table}[h!]
  \begin{center}
    \caption{\emph{\textbf{Maximum Masses and the Corresponding Radii of Static NS for the $a$-attractors  and for GR}}}
    \label{table1}
    \begin{tabular}{|r|r|r|r|}
     \hline
      \textbf{Model}   & \textbf{APR EoS} & \textbf{SLy EoS} & \textbf{WFF1 EoS}
      \\  \hline
      \textbf{GR} & $M_{max}= 2.18739372\, M_{\odot}$ & $M_{max}= 2.04785291\, M_{\odot}$ & $M_{max}= 2.12603999\, M_{\odot}$
      \\  \hline
      \textbf{$a$-attractors $a=0.6$} & $M_{max}= 2.42368632\,M_{\odot}$ & $M_{max}= 2.32152569\,M_{\odot}$
      &$M_{max}= 2.27222814\, M_{\odot}$ \\  \hline
       \textbf{$a$-attractors Radii $a=0.6$} & $R= 10.89673795$km &
$R= 10.81663975$km
      &$R= 9.8993922$km \\  \hline
      \textbf{$a$-attractors large $a$} & $M_{max}=2.41736711\,M_{\odot}$ & $M_{max}= 2.24807309\,M_{\odot}$ &$M_{max}= 2.34266656\, M_{\odot}$ \\  \hline
\textbf{$a$-attractors Radii large $a$} & $R= 10.35829055$km & $R=
9.98770954$km
      &$R= 9.28095863$km \\  \hline
    \end{tabular}
  \end{center}
\end{table}
In order to quantify the effect brought along by the different
values of the parameter $a$, in Fig. \ref{plot2} we present the
same $M-R$ graphs corresponding to distinct values of the
parameter $a$ for the three different EoSs we used. The left plot
corresponds to $a=0.6$ while the right plot corresponds to
$a=1000$ for the WFF1 EoS and $a=10^4$ for the APR and SLy EoS.
From Fig. \ref{plot2} one can see the effect of decreasing the
value of $a$, which is the increase of the radius of static NSs in
hydrodynamic equilibrium.

Before closing we need to mention another class of $a$-attractor
potentials which yield exactly the same theory in the Einstein
frame, with the same conformal factor. The Jordan frame action in
Geometrized units for these alternative $a$-attractors is,
\begin{equation}\label{ta}
\mathcal{S}=\int
d^4x\frac{\sqrt{-g}}{16\pi}\Big{[}f(\phi)R-\omega(\phi)g^{\mu
\nu}\partial_{\mu}\phi\partial_{\nu}\phi-U_J(\phi)\Big{]}+S_m(\psi_m,g_{\mu
\nu})\, ,
\end{equation}
with the potential $U(\phi)$ being equal to,
\begin{equation}\label{alphaextra}
U_J=\mathcal{U}_0\left(1-f(\phi )\right)^2\, .
\end{equation}
Choosing again the kinetic coupling function $\omega(\phi)$ as in
Eq. (\ref{omegaphi}), the function $f(\phi)$ as a function of the
canonical Einstein frame scalar field is given again by Eq.
(\ref{fvarphi}), while the resulting Einstein frame potential is
given again by the potential (\ref{einsteinframepotentialfinal})
for $n=1$. In addition, our numerical analysis indicated that by
choosing a different number $n$ in the potential
(\ref{einsteinframepotentialfinal}), does not change significantly
the resulting NS phenomenology.

In view of our results, a strong question is whether the
inflationary attractor behavior continues to NSs phenomenology
too. This question is difficult to answer for the moment, because
we need more data and analysis of several inflationary attractor
potentials. However, one is certain, that a whole class of
different arbitrary types of non-minimal couplings in the Jordan
frame, with three different classes of potentials yield similar
results.

\section*{Concluding Remarks}

In this work we investigated the implications of a class of viable
inflationary scalar potentials on NS. Specifically we were
interested on $a$-attractor potentials which are known to provide
not only a viable inflationary era, but also predict the same form
for the spectral index of the primordial curvature perturbations
and for the tensor-to-scalar ratio, regardless the functional form
of the non-minimal coupling in the Jordan frame. The $a$-attractor
potentials contain a parameter $a$, which when it takes values
near unity or much smaller than unity, the resulting inflationary
phenomenology is viable. On the contrary when the parameter $a$
takes large values, the inflationary phenomenology is not viable.
After presenting the mechanism of obtaining the $a$-attractor
potentials in the Einstein frame, starting from a class of Jordan
frame models, we calculated the TOV equations for a spherically
symmetric and static metric in the Einstein frame, having in mind
that for the results we need to convert the Einstein frame
quantities to their Jordan frame counterparts. With regard to the
EoS, we used a piecewise polytropic EoS, with the low density part
being the WFF1 or the APR or the SLy EoS. Accordingly, we
numerically solved the TOV equations using a python 3 based
numerical code, and we derived the $M-R$ graphs for the
$a$-attractor theory. The resulting picture was quite interesting,
since for all the studied cases, when small central densities and
large $a$ vales are considered, the $M-R$ graphs of GR and of the
$a$-attractor models coincide, however for larger central
densities, the two graphs become clearly different. It is
interesting to recall that the inflationary theory for large
values of $a$ is not viable. Now, when small values of $a$ are
considered, $a$-attractor $M-R$ diagrams deviated significantly
from the corresponding GR graphs. Interestingly enough for small
values of $a$, the inflationary theory is viable and compatible
with the latest Planck observational constraints on inflation.
Moreover, in all the cases, except for the WFF1 case for large
$a$, the radii of the NSs for the $a$-attractor models satisfy
both the GW170817 constraints derived in Ref.
\cite{Bauswein:2017vtn}, namely that the radius of a $M\sim 1.6
M_{\odot}$ NS must be larger than $R=10.68^{+15}_{-0.04}$km, and
the radii of the NSs corresponding to the maximum mass
configurations, must be larger than $R=9.6^{+0.14}_{-0.03}$km.
These constraints are violated for the WFF1 EoS, only for large
$a$, since the theory mimics basically GR for those values of $a$.
Before closing we need to mention that the $a$-attractor $M-R$
graphs seem to interpolate between the GR and the small $a$
$a$-attractors, for intermediate central densities however.
Interestingly enough, the large $a$ potentials are essentially
$\sim \varphi^2$ potentials. An important issue we did not address
is whether the observed attractor behavior in the inflationary
theories, continues to the level of NS, quantified differently of
course. It seems that our preliminary results indicate deep and
not yet completely understood connections between non-minimal
inflationary attractors potentials and NSs phenomenology in
scalar-tensor theory. In fact the inflationary attractor behavior
might or might not be observed in NSs phenomenology, and in both
cases the result will be interesting. In detail, if the attractor
behavior is observed, this will be a direct correspondence between
the inflationary and NS phenomenological theories. In the case
that the attractor behavior is not observed, this will be a direct
way of discriminating inflationary attractors at the NS level. For
the moment we do not have enough data to address this important
issue, and we hope to address this in a future work. Another issue
that is worth discussing, is the Einstein-Gauss-Bonnet extension
of the present theories
\cite{Hwang:2005hb,Kanti:2015pda,DeLaurentis:2015fea,Yi:2018dhl,Kleihaus:2019rbg,Bakopoulos:2019tvc},
which is string theory motivated. This study can be done for both
the minimal and non-minimal couplings of the scalar field to the
Ricci scalar, since there is no correspondence between the minimal
and non-minimal couplings when the Gauss-Bonnet coupling is
present. Thus only the Jordan frame theory can be studied. We hope
to address some of these issues in a future work.

\section*{Acknowledgments}

V.K. Oikonomou is indebted to N. Stergioulas and his MSc student
Vaggelis Smyrniotis for the many hours spend on neutron star
physics discussions and for sharing his professional knowledge on
numerical integration of neutron stars in python. This work was
supported by MINECO (Spain), project PID2019-104397GB-I00 and
PHAROS COST Action (CA16214) (SDO).

\end{document}